\documentclass[10pt, conference, letterpaper]{IEEEtran}
\IEEEoverridecommandlockouts
\usepackage{cite}
\usepackage{marvosym} 
\usepackage{amsmath,amssymb,amsfonts}
\usepackage{algorithmicx}   
\usepackage{algpseudocode}  
\usepackage{algorithm}      
\algnewcommand{\Initialization}{\item[\textbf{Initialization:}]}
\algnewcommand{\Input}{\item[\textbf{Input:}]}
\algnewcommand{\Output}{\item[\textbf{Output:}]}

\usepackage{graphicx}
\usepackage{subcaption}  
\usepackage{textcomp}
\usepackage{xcolor}
\def\BibTeX{{\rm B\kern-.05em{\sc i\kern-.025em b}\kern-.08em
    T\kern-.1667em\lower.7ex\hbox{E}\kern-.125emX}}

\DeclareRobustCommand*{\IEEEauthorrefmark}[1]{%
    \raisebox{0pt}[0pt][0pt]{\textsuperscript{\footnotesize\ensuremath{#1}}}}

\begin{document}

\title{User-Intent-Driven Semantic Communication via Adaptive Deep Understanding\\
\thanks{\textsuperscript{*}\textit{Equal contribution: Peigen Ye, Jingpu Duan}; 

\textsuperscript{\Letter}\textit{Corresponding authors: Hongyang Du, Yulan Guo.}}
}


\author{\IEEEauthorblockN{
Peigen Ye\IEEEauthorrefmark{1, 2, 3},\ 
Jingpu Duan\IEEEauthorrefmark{2},\ 
Hongyang Du\IEEEauthorrefmark{3},\ 
Yulan Guo\IEEEauthorrefmark{1}
}

\IEEEauthorblockA{\IEEEauthorrefmark{1}Sun Yat-sen University, Shenzhen, China}
\IEEEauthorblockA{\IEEEauthorrefmark{2}Pengcheng Laboratory, Shenzhen, China}

\IEEEauthorblockA{\IEEEauthorrefmark{3}The University of Hong Kong, Hong Kong SAR, China}

}

\maketitle

\begin{abstract}
Semantic communication focuses on transmitting task-relevant semantic information,
aiming for intent-oriented communication. While existing systems improve efficiency by extracting key semantics, they still fail to deeply understand and generalize users' real intentions. 
To overcome this, we propose a user-intention-driven semantic communication system that interprets diverse abstract intents. First, we integrate a multi-modal large model as semantic knowledge base to generate user-intention prior. Next, a mask-guided attention module is proposed to effectively highlight critical semantic regions. Further, a channel state awareness module ensures adaptive, robust transmission across varying channel conditions. 
Extensive experiments demonstrate that our system achieves deep intent understanding and outperforms DeepJSCC, e.g., under a Rayleigh channel at an SNR of 5 dB, it achieves improvements of 8\%, 6\%, and 19\% in PSNR, SSIM, and LPIPS, respectively.
\end{abstract}


\section{Introduction}


Semantic Communication (SemCom), as a novel and promising paradigm, differs from conventional communication systems that prioritize error-free symbol transmission. Instead, it emphasizes transmitting task-relevant semantic content. By extracting and preserving essential semantic elements, SemCom effectively reduces transmission overhead, thereby significantly improving both efficiency and robustness.

Fundamentally, the ultimate goal of information transmission is to convey the user's real intent. Thus, the ideal form of SemCom is intent-oriented communication~\cite{zhang2022toward}.
Several studies have explored this objective. Weng \textit{et al.} \cite{weng2023deep} focused on transmitting semantic information of speech, where at the receiver, they synthesized the user's voice by combining personalized voice features learned from the user with the decoded text, thereby reducing information redundancy while preserving the intent of reproducing the user's voice.
Zhong \textit{et al.} \cite{Zhong10901654} proposed a method focusing on the transmission of generation seeds to reduce transmission size, enhancing image quality under low SNR.
Ye \textit{et al.}~\cite{ye2024codebook}
and Zhou \textit{et al.} \cite{10901060} 
focused on visual perceptual quality in image transmission rather than pixel-level precision, designing a generative SemCom system based on a codebook Semantic Knowledge Base (SKB), which significantly enhanced the visual fidelity of the received images.
With the advancement of Large Models (LMs), several studies~\cite{jiang10670195} have explored their use as SKBs to facilitate the extraction of core semantic content. Guo \textit{et al.} \cite{Guo10843783} leveraged LMs to transform images into natural language descriptions containing key semantic information for semantic transmission, thereby reducing communication overhead.
Zhao \textit{et al.} \cite{zhao10531769} employed a Large Language Model (LLM) to generate textual descriptions of images and then transmitted fused features of both image and text to enable robust transmission.
These studies, to varying degrees, have achieved effective extraction and transmission of core semantic contents, thereby improving communication efficiency.

\begin{figure}[t]  
  \centering
  \includegraphics[width=3.32in]{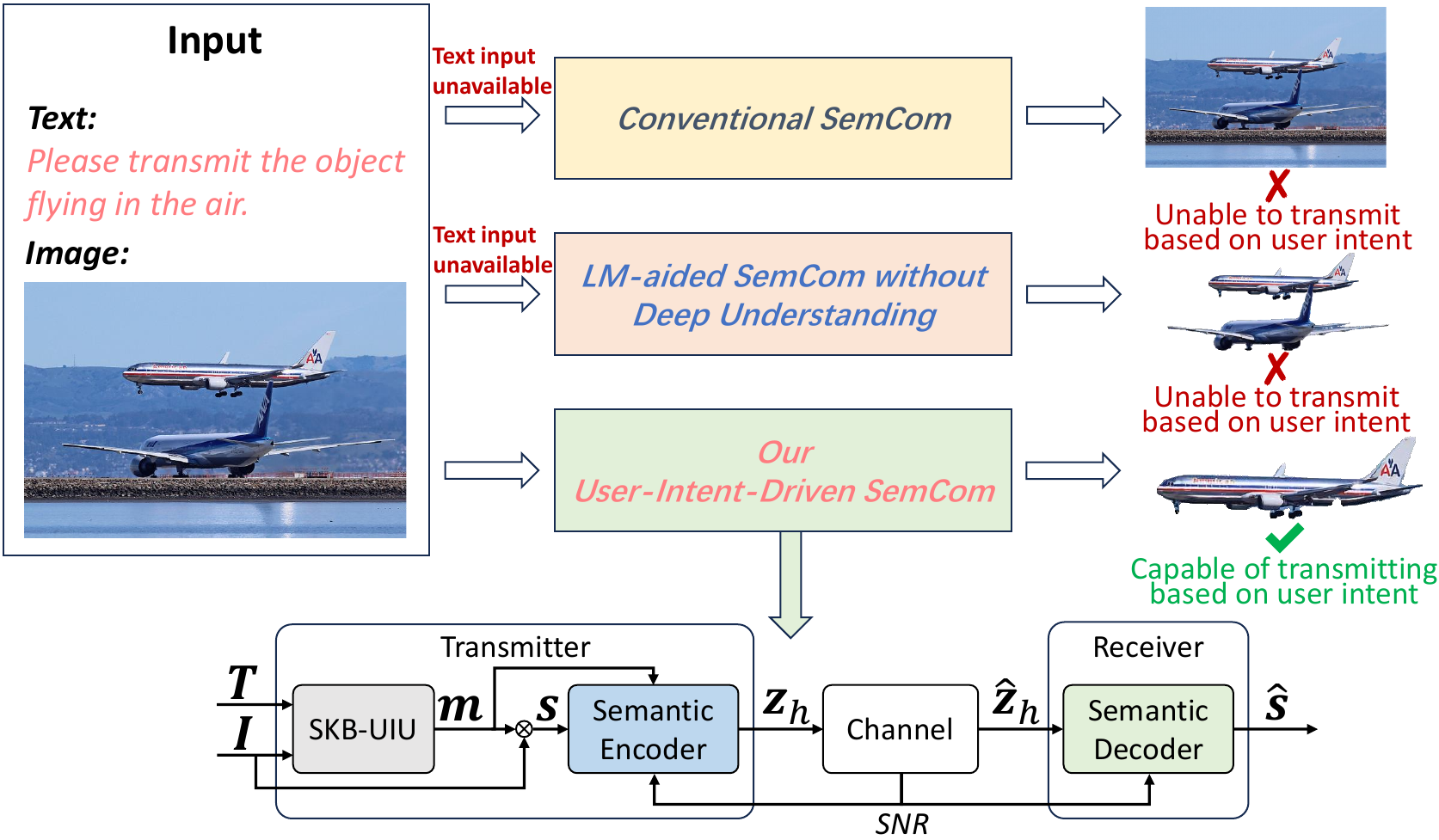}
  \caption{A comparison of various SemCom systems and the framework of our User-Intent-Driven SemCom system.}
  \label{fig1:system}
\end{figure}

However, most current SemCom systems --- including those assisted by LMs --- still rely on SKBs constructed from statistical experience to determine core semantic content, lacking deep understanding and adaptive reasoning capabilities regarding the users’ real intent in specific communication tasks. Such systems are typically tailored to specific types of communication tasks and struggle to adapt to the diverse intent scenarios of different users, thereby exhibiting limited generalizability. 
For example, the focal points of \textit{User A} and \textit{User B} may differ significantly, and models or knowledge bases trained based on \textit{User A}'s preferences are unlikely to transfer to the needs of \textit{User B}. 
To achieve intent-oriented communication across different users, existing systems generally require training separate models or personalized knowledge bases. Further, even for the same user, their intent in different communication tasks may deviate from their previous behavioral patterns. 
A more challenging issue arises because users' real intent is often abstract or implicit in their natural language expressions, which existing communication systems may struggle to accurately understand. For instance, the user instruction ``\emph{Please transmit the object flying in the air}" poses a semantic challenge for traditional systems, which may fail to understand the meaning of ``\emph{the object flying in the air}".
To sum up, constructing a SKB endowed with strong knowledge representation and reasoning capabilities is key to realizing real intent-oriented communication. 

To overcome the inability of current semantic communication systems to understand user intent,
we innovatively position a multi-modal large model as SKB for deep understanding user intent, leveraging their powerful capacity to interpret users' explicit or implicit instructions, identify users' real intent, and generate Region-of-Interest (ROI) masks. This approach addresses the limitations of traditional SKBs in terms of understanding and generalization.
To further enhance the model's attention to the user's ROIs, we design a mask-guided attention module that strengthens the perception of key areas \cite{li2022contextual}. In addition, to enable the model to proactively perceive the channel conditions, we incorporate a channel state awareness module by embedding noise distribution into the feature representations \cite{zhang2018ffdnet}. 
Main contributions of this paper lie in the following aspects. 
\begin{itemize}
\item We introduce a multi-modal large model as the SKB for deep understanding, 
overcoming the limitations of traditional knowledge bases in understanding and generalization.
\item A mask-guided attention module is proposed to enhance the ability of existing systems to focus on critical regions.
\item A channel state awareness module is incorporated to enable proactive sensing of channel conditions, improving the system's stability and adaptability under varying levels of channel noise.
\item By integrating these modules, we construct a User-Intent-Driven SemCom (UIDSC) system capable of deeply understanding users' real intentions and adapting to physical channel, achieving intent-driven, adaptive, and robust transmission. The effectiveness of the proposed system is validated through extensive experiments.
\end{itemize}

\section{System Model}

This section introduces the fundamental framework of the proposed SemCom system. As illustrated in Fig.~\ref{fig1:system}, the system consists of three main components: the transmitter, the physical channel, and the receiver. The transmitter includes a Semantic Knowledge Base for User-Intent-Understanding (SKB-UIU) and a semantic encoder, while the receiver incorporates a semantic decoder. The communication process is described as follows: the input image and the user's intention in natural language form are provided to the transmitter; the transmitter, by deep understanding the user's intention (which may be implicit), extracts the ROIs from the input image and encodes them as a feature vector for transmission; the feature vector is then transmitted through the physical channel to the receiver; the receiver decodes the feature vector and reconstructs a high-quality image of the user’s ROIs.


\begin{figure*}[t]  
  \centering
  \includegraphics[width=\textwidth]{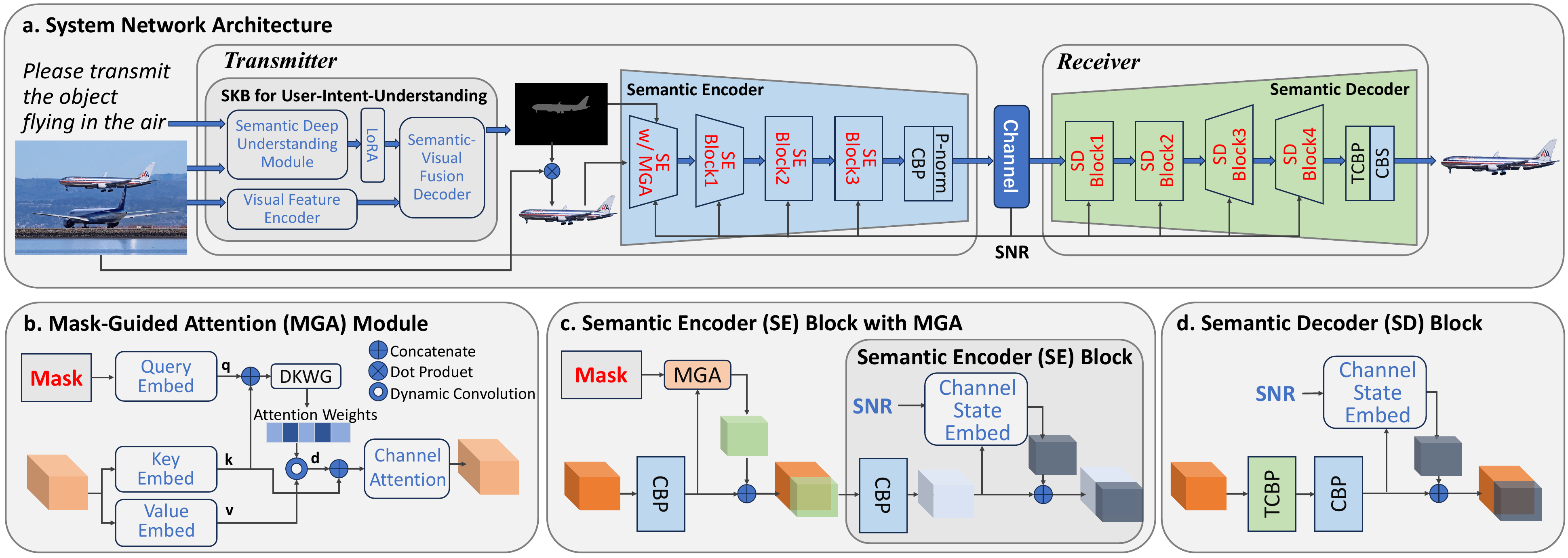}
  \caption{The network architecture of the proposed SemCom system.
  CBP refers to the Conv-BN-PReLU block; CBS refers to the Conv-BN-Sigmoid block; TCBP refers to the TConv-BN-PReLU block; P-norm denotes the power normalization operation.}
  \label{fig:two-stage network}
\end{figure*}

\subsection{Transmitter}

As shown in Fig.~\ref{fig1:system}, the input image $\mathbf{I} \in {\mathbb{R} ^{H \times W \times 3}}$ and the user's intention $\mathbf{T} \in {\mathbb{R} ^{n \times d}}$ are inputted into the SKB-UIU. The SKB-UIU, through a deep interpretation of the user's intention, extracts the ROI from the input image and outputs a mask $\mathbf{m} \in {\mathbb{R} ^{H \times W \times 1}}$. This process can be expressed as follows:

\begin{equation}
\mathbf{m} =\mathbf{UIU}_{\psi} \left ( \mathbf{T} , \mathbf{I}  \right ) {,}
\end{equation}
where $\mathbf{UIU}_{\psi}\left ( \cdot  \right )$ represents the SKB-UIU module, which consists of a large-scale vision understanding model and a multi-modal LLM, with its parameters denoted as ${\psi}$.

The extracted ROI mask $\mathbf{m}$ is multiplied with the original image $\mathbf{I}$ to obtain the ROI image $\mathbf{s}$. The image $\mathbf{s}$ and the mask $\mathbf{m}$ are then fed into the channel-aware Semantic Encoder ($\mathbf{SE}$) to generate a feature vector, as follows:
\begin{equation}
\mathbf{z} _{h}=\mathbf{SE} _{\theta } \left ( \mathit{\mathbf{s}}  , \mathbf{m} , \mathit{snr} \right ) {,}
\end{equation}
where $\mathbf{SE}_{\theta}\left ( \cdot  \right )$ represents the semantic encoder function with respect to parameters ${\theta }$. It consists of an attention module, a channel-aware module, and a series of convolutional blocks. 
Each SE block is aware of the SNR and can be configured with a desired downsampling factor.


\subsection{Physical Communication Channel}

Assuming a single communication link for image SemCom, we employ the Additive White Gaussian Noise (AWGN) channel model and the Rayleigh slow fading channel model, 
which are widely used in Wave-based Wireless Transmission,
as the physical channel model~\cite{bourtsoulatze2019deep}. 
The output feature vector of the semantic encoder is organized into a complex signal $\mathbf{z} _{h}$, which is then transmitted through the physical channel to the receiver, producing the complex signal $\hat{\mathbf{z}}_{h}$ at the receiver. 

i) For the AWGN channel, the process can be expressed as,
${\hat{\mathbf{z}}_h} = {\mathbf{z}_h} + \mathbf{n}$,
where $\mathbf{n}\in\mathbb{C}^{k}$ consists of independent and identically distributed (i.i.d.) with zero mean and variance $\sigma^{2}$, i.e., $\mathbf{n}\sim \mathcal{CN} \left ( 0, \sigma ^{2} \mathbf{I} \right ) $.

ii) For the Rayleigh channel, the process can be expressed as,
${\hat{\mathbf{z}}_h} = {\mathbf{h}\mathbf{z}_h} + \mathbf{n}$,
where the channel gain $\mathbf{h}\sim \mathcal{CN} \left ( 0, H_{c}\right )$ is a complex Gaussian random variable,
and $\mathbf{n}\sim \mathcal{CN} \left ( 0, \sigma ^{2} \mathbf{I} \right ) $. We set $H_{c}= 1$ according to previous works~\cite{bourtsoulatze2019deep}.

\subsection{Receiver}

The receiver consists of a series of Semantic Decoder (SD) blocks with channel state awareness, along with convolutional blocks. The signal, transmitted through the wireless physical channel, is received by the receiver and then fed into the semantic decoder, which decodes a high-quality ROI image for the user. This process can be expressed as follows:

\begin{equation}
\hat{\mathbf{s} } =\mathbf{SD} _{\zeta } \left ( \hat{\mathbf{z}}_{h}, \mathit{snr}\right ) {,}
\end{equation}
where ${\mathbf{SD}_\mathbf{\zeta} }\left( \cdot \right)$ represents the semantic decoder function with respect to parameters $\mathbf{\zeta}$. Each SD block is aware of the SNR and can be configured with a desired upsampling factor.

\section{The Proposed Method}
\label{sec:third_section}
In previous research on SemCom, enabling communication systems to deeply understand user's intention has been a significant challenge. However, thanks to the development of large-scale artificial intelligence models, network models are now more capable than ever of understanding the intentions embedded in human natural language. Inspired by related works, we propose an innovative SemCom system with user's intention as the transmission target. As shown in Fig. \ref{fig:two-stage network}, the proposed system consists of several key modules: SKB for user-intent-understanding, mask-guided attention module, channel state embedding module, semantic encoder, and semantic decoder. These modules are described in detail below.

\subsection{SKB for User-Intent-Understanding}\label{SKB}

Inspired by \cite{lai2024lisa}, the SKB-UIU module primarily operates based on pre-trained LMs to deeply understand user's intention. First, the semantic deep understanding module, which is a multi-modal LLM \cite{liu2023visual}, performs a deep interpretation of the input image based on the text instruction carrying the user’s intention. Leveraging the powerful semantic reasoning capabilities of the LLM, it infers the user’s intent (either explicit or implicit) and then generates the semantic embedding vector of the ROI. Meanwhile, the visual feature encoder, based on the Vision Transformer-Huge architecture \cite{kirillov2023segment}, extracts multi-level visual features from the image and constructs a high-dimensional semantic feature map. The semantic-visual fusion decoder then integrates the semantic embeddings and visual features through a dynamic interaction mechanism, ultimately decoding a pixel-level binary mask. The module employs Low-rank Adaptation (LoRA) \cite{hu2022lora} technology for parameter-efficient fine-tuning of the LLM parameters, optimizing only the local weights to avoid degradation of language capabilities.

\subsection{Mask-Guided Attention}\label{MGA}

We propose a Mask-Guided Attention (MGA) module, which adaptively enhances the features of the target region by introducing a semantic mask as an explicit guide. As shown in Fig. \ref{fig:two-stage network}-b, the module first employs a convolutional network to encode the spatial context of the binary mask, generating an attention query vector $\mathbf{q}$ aligned with the feature map channels. Next, the attention query vector $\mathbf{q}$ is concatenated with the static context features (i.e., the encoded key vector $\mathbf{k}$), and passed through a Dynamic Kernel Weight Generator (DKWG), a lightweight convolutional network, to dynamically generate spatially adaptive convolutional weights. The dynamic convolution is then employed to capture detailed features of the target region. Finally, a dual-path feature fusion mechanism is used to combine channel attention, performing a weighted aggregation of the dynamic features $\mathbf{d}$ and static features $\mathbf{k}$.

\subsection{Channel State Embedding}\label{CSE}

The Channel State Embedding (CSE) module maps the Signal-to-Noise Ratio (SNR) to the noise distribution in the feature space, assisting the model in perceiving the channel state. First, the average signal power $P _{s} $ is calculated based on the input features $\mathbf{z}_{i}\in \mathbb{R} ^{h\times w\times c} $, $P_{s}=\frac{1}{K} {\textstyle \sum_{i=1}^{K}}\mathbf{z}_{i}^{2}$,
%
where $K=h\times w\times c$.
Then, the noise level $\sigma _{n}^{2}$ is calculated based on the SNR feedback from the channel, $\sigma _{n}^{2} =\frac{P_{s}  }{10^{\mathit{snr} /10} }$.


\begin{figure}[t]  
  \centering
  \includegraphics[width=3.3in]{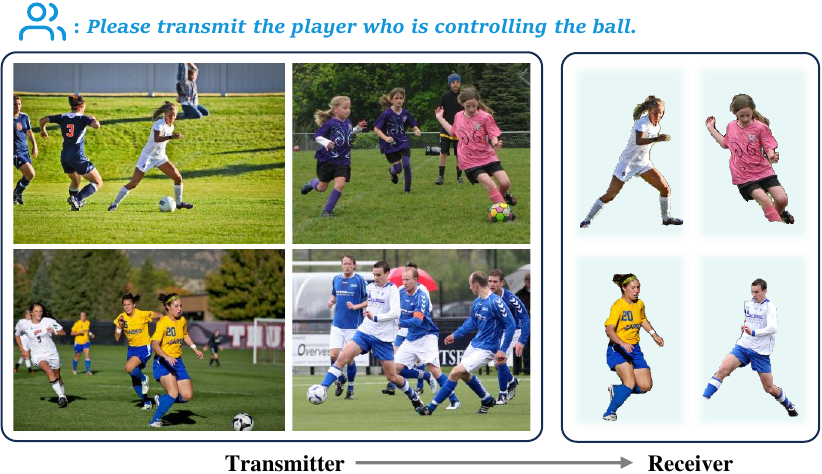}
  \caption{A transmission example in the User-Intent-Driven SemCom system. The input instruction is ``\textit{Please\ transmit\ the\ player\ who\ is\ controlling\ the\ ball}."}
  \label{fig:Pic3_intention}
\end{figure}

The noise map $\mathbf{M}_{n}$ is obtained by extending $\sigma _{n}^{2}$ to match the size of $\mathbf{z}_{i}$, and is then fused with the input features through channel concatenation. The noise map generated by the CSE is embedded at each feature layer of the semantic encoder and semantic decoder, enabling the model to perceive the channel state at multiple levels.

\begin{algorithm}[htbp]
\caption{Two-stage Training Strategy for UIDSC System}
\label{alg:training}
\begin{algorithmic}[1]
\Statex //Stage1:
\Initialization{Initialize model without MGA structure and parameters $\theta^{(0)}, \zeta^{(0)}$, $i = 0$}
\Input{A batch of ORI-train images $\mathbf{s}$, channel state $\mathit{snr}$}
\While{end criterion is not met}
    \State $\mathbf{z} _{h} \leftarrow \mathbf{SE} _{\theta }^{\mathrm{w/o\ MGA}} \left ( \mathit{\mathbf{s}} , \mathit{snr} \right )$
    \State Transmit $\mathbf{z} _{h}$ over wireless channel
    \State $\hat{\mathbf{s} }\leftarrow\mathbf{SD} _{\zeta } \left ( \hat{\mathbf{z}}_{h}, \mathit{snr}\right )$
    \State Compute loss $\mathcal{L}_{\text{L2}}(\theta, \zeta)=\mathrm{MSE} \left ( \mathbf{s} ,\hat{\mathbf{s}}   \right ) $
    \State Update parameters via SGD:
    \begin{align*}
    \theta^{(i+1)} &\leftarrow \theta^{(i)} -  \nabla_{\theta^{(i)}} \mathcal{L}_{\text{L2}}\left (\theta ,\zeta  \right ) \quad \\
    \zeta^{(i+1)} &\leftarrow \zeta^{(i)} -  \nabla_{\zeta^{(i)}} \mathcal{L}_{\text{L2}}\left (\theta ,\zeta  \right ) \quad
    \end{align*}
    \State $i \leftarrow i + 1$
\EndWhile
\Output{SE and SD parameters $\theta^{Stage1}, \zeta^{Stage1}$}
\Statex //Stage2:
\Initialization{Load model with MGA structure and parameters $\theta^{(0)}=\theta^{Stage1}, \zeta^{(0)}=\zeta^{Stage1}$, $i = 0$}
\Input{A batch of SEG-train images $\mathbf{s}$, MASK-train images $\mathbf{m}$, channel state $\mathit{snr}$}
\While{end criterion is not met}
    \State $\mathbf{z} _{h} \leftarrow \mathbf{SE} _{\theta }^{\mathrm{w/\ MGA}} \left ( \mathit{\mathbf{s}}  , \mathbf{m} , \mathit{snr} \right )$
    \State Transmit $\mathbf{z} _{h}$ over wireless channel
    \State $\hat{\mathbf{s} }\leftarrow\mathbf{SD} _{\zeta } \left ( \hat{\mathbf{z}}_{h}, \mathit{snr}\right )$
    \State Compute loss $\mathcal{L}_{\text{L2}}(\theta, \zeta)=\mathrm{MSE} \left ( \mathbf{s} ,\hat{\mathbf{s}}   \right ) $
    \State Update parameters via SGD:
    \begin{align*}
    \theta^{(i+1)} &\leftarrow \theta^{(i)} -  \nabla_{\theta^{(i)}} \mathcal{L}_{\text{L2}}\left (\theta ,\zeta  \right ) \quad \\
    \zeta^{(i+1)} &\leftarrow \zeta^{(i)} -  \nabla_{\zeta^{(i)}} \mathcal{L}_{\text{L2}}\left (\theta ,\zeta  \right ) \quad
    \end{align*}
    \State $i \leftarrow i + 1$
\EndWhile
\Output{SE and SD parameters $\theta^{final}, \zeta^{final}$}
\end{algorithmic}
\end{algorithm}

\addtolength{\topmargin}{0.05in}

\subsection{Semantic Encoder}\label{SE}

The semantic encoder consists of the MGA module, CSE module, power normalization layer, and other components, forming a deep feature encoding architecture designed for noise robustness. The encoder enhances semantic representation through a multi-stage feature encoding pipeline: First, the attention fusion unit, based on the MGA module, focuses on the target region using the guidance of the semantic mask; next, the CSE module transforms the SNR parameter into a noise map, which is injected into the feature space via channel concatenation, enabling the encoding process to perceive the channel noise distribution explicitly; Finally, a multi-stage downsampling structure progressively fuses semantic features with noise priors, and then the power normalization layer constrains the output feature power distribution.

The encoder demonstrates significant advantages under low SNR conditions, effectively suppressing channel noise interference through the collaborative optimization of attention mechanisms and channel perception, providing the receiver with semantic-rich and channel-adaptive encoded features.

\begin{figure*}[ht]
  \centering
  \begin{subfigure}{0.29\textwidth}
    \includegraphics[width=\linewidth,trim=20 10 30 10,clip]{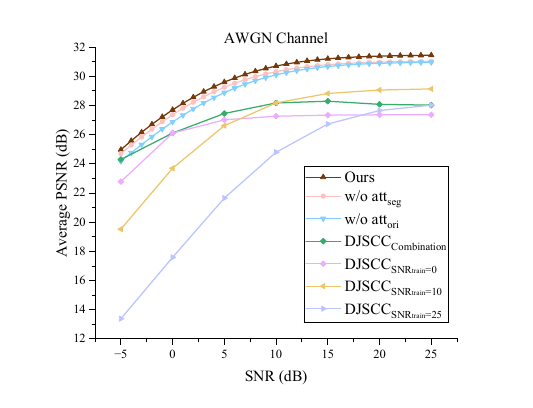}
    \caption{SNR-PSNR in AWGN Channel}
    \label{fig:subfig1}
  \end{subfigure}
  \hfill
  \begin{subfigure}{0.29\textwidth}
    \includegraphics[width=\linewidth,trim=20 10 30 10,clip]{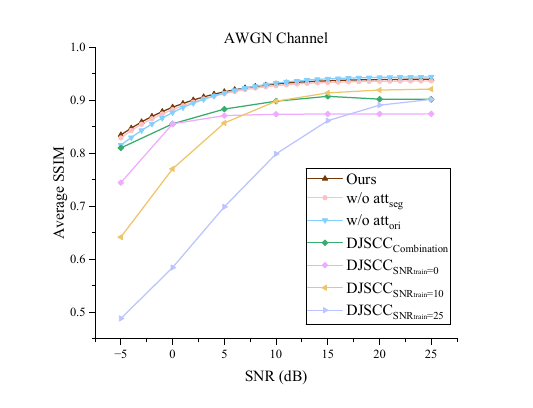}
    \caption{SNR-SSIM in AWGN Channel}
    \label{fig:subfig2}
  \end{subfigure}
  \hfill
  \begin{subfigure}{0.29\textwidth}
    \includegraphics[width=\linewidth,trim=20 10 30 10,clip]{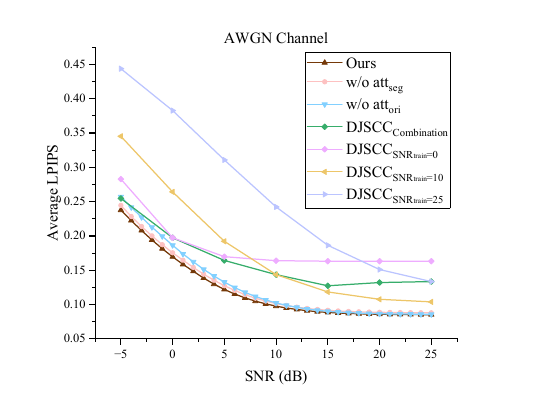}
    \caption{SNR-LPIPS in AWGN Channel}
    \label{fig:subfig3}
  \end{subfigure}

  \begin{subfigure}{0.29\textwidth}
    \includegraphics[width=\linewidth,trim=20 10 30 10,clip]{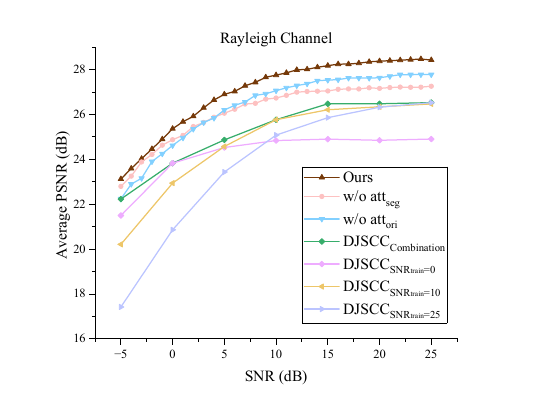}
    \caption{SNR-PSNR in Rayleigh Channel}
    \label{fig:subfig4}
  \end{subfigure}
  \hfill
  \begin{subfigure}{0.29\textwidth}
    \includegraphics[width=\linewidth,trim=20 10 30 10,clip]{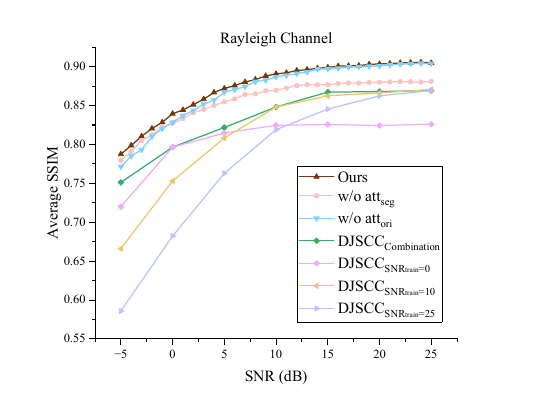}
    \caption{SNR-SSIM in Rayleigh Channel}
    \label{fig:subfig5}
  \end{subfigure}
  \hfill
  \begin{subfigure}{0.29\textwidth}
    \includegraphics[width=\linewidth,trim=20 10 30 10,clip]{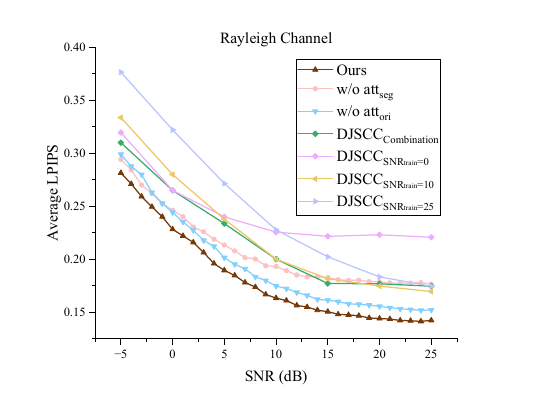}
    \caption{SNR-LPIPS in Rayleigh Channel}
    \label{fig:subfig6}
  \end{subfigure}

  \caption{The performance metrics, including PSNR $\left ( \uparrow  \right )$, SSIM $\left ( \uparrow  \right )$, and LPIPS $\left ( \downarrow   \right )$, for different methods across the SNR range of $-5$ to $25$ dB. \textbf{Ours} refers to the complete our method; \textbf{w/o\ att$\mathrm{_{seg}}$} refers to our method without the MGA module, trained on SEG-train; \textbf{w/o\ att$\mathrm{_{ori}}$} refers to our method without the MGA module, trained on ORI-train; \textbf{DJSCC$\mathrm{_{SNR_{train}=\left \{0,10,25\right \}}}$ } represents models trained under different channel conditions with $\mathrm{SNR_{train}}$ set to $0$, $10$, and $25$ dB, respectively; \textbf{DJSCC$\mathrm{_{Combination} } $} represents the combination curve of the models trained independently at various $\mathrm{SNR_{train}}$ levels (e.g., the value on the \textbf{DJSCC$\mathrm{_{Combination} } $} curve corresponding to $\mathrm{SNR=10}$ dB is derived from the \textbf{DJSCC$\mathrm{_{SNR_{train}=10} } $} model).}
  \label{fig:4-metrics}
\end{figure*}

\subsection{Semantic Decoder}\label{SD}

The Semantic Decoder consists primarily of semantic decoder blocks (as shown in Fig. \ref{fig:two-stage network}-d) and serves as a feature reconstruction framework with channel state awareness. As depicted in Fig. \ref{fig:two-stage network}-a, the decoder reconstructs semantic features through a multi-stage upsampling structure: First, in each semantic decoder block, the CSE module is introduced to map the SNR parameter to a noise map, which is injected into the decoding process via channel concatenation, ensuring the feature reconstruction is adaptive to the channel noise distribution; next, a composite unit of transposed and regular convolutions is used to enhance spatial resolution and refine local features; finally, a Sigmoid activation function constrains the output range, producing a semantic reconstruction result that aligns with the pixel value domain. 

The decoder forms a symmetric encoding-decoding structure with the semantic encoder, achieving end-to-end channel state alignment through the multi-stage embedding of the CSE module, ensuring the alignment between the encoding feature process and the decoding reconstruction process.


\section{Experiment and Evaluation}
\label{IV-experiment}

\subsection{Datasets and Settings}

\textbf{The training dataset:} We constructed three training subsets based on the COCO 2017 \cite{lin2015microsoftcococommonobjects} dataset: the original image training dataset (ORI-train), the segmented image training dataset (SEG-train), and the mask image training dataset (MASK-train). ORI-train is the original training dataset provided by COCO. SEG-train and MASK-train are obtained through preprocessing of the instance segmentation annotations and corresponding images from the COCO dataset, without altering the original image quality.
During training, input images are randomly cropped or padded to a fixed resolution of $256 \times 256$ to support batch training. During testing, the original resolution of the input images is preserved.

\textbf{The test dataset:} Since the COCO dataset does not publicly release annotations for its test dataset, we randomly selected 1,000 images from the COCO validation dataset, ensuring no overlap with the training and validation datasets of the current experiment. These images and their corresponding instance segmentation annotations are used to construct the segmented image test dataset (SEG-test).

\textbf{Settings:} In our system, the SKB-UIU module uses pre-trained weights from \cite{lai2024lisa}. The subsequent modules follow a two-stage training strategy, as outlined in Algorithm \ref{alg:training}: Stage 1 is trained on the ORI-train dataset to ensure the model without MGA has good generalization ability; Stage 2 is trained on the SEG-train and MASK-train datasets to ensure the model’s focus on the target region. Both stages use the Adam optimizer with an initial learning rate of $1e-4$, a batch size of $32$, and $100$ epochs. The loss function is the L2 loss between $\mathbf{s}$ and $\hat{\mathbf{s}}$. This method is implemented based on the PyTorch framework and trained using one
GeForce RTX 4090 GPU.

\subsection{Simulation and Evaluation}

This section systematically validates the performance advantages of the proposed method: First, it verifies the SemCom capability based on user intention; then, a comparative analysis of transmission quality is conducted under diverse channel conditions, comparing the proposed method with baseline approaches, including DeepJSCC \cite{bourtsoulatze2019deep} and its variants.

\begin{figure*}[t]  
  \centering
  \includegraphics[width=6.9in]{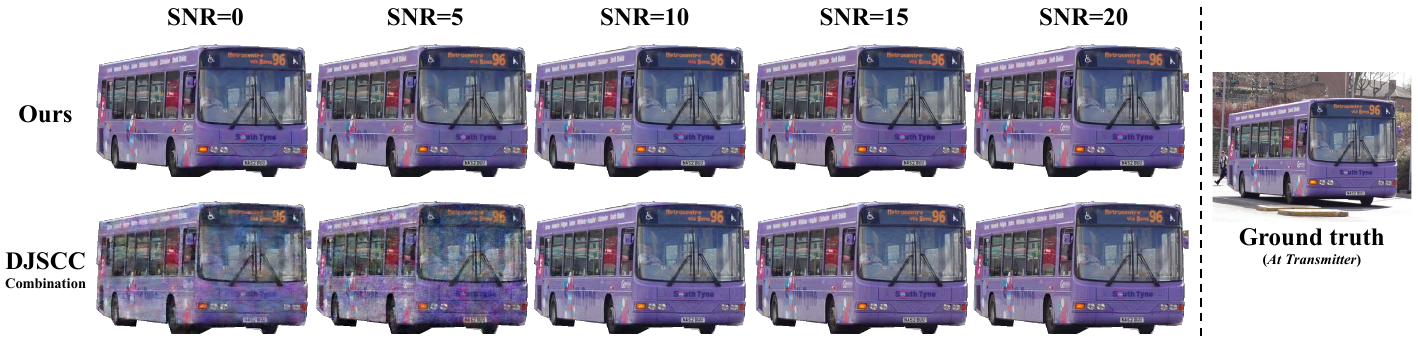}
  \caption{The output results of image transmission for different methods under various SNR levels (0, 5, 10, 15, and 20 dB) over the Rayleigh channel.}
  \label{fig:Pic5_outputPicShow}
\end{figure*}

Fig. \ref{fig:Pic3_intention} illustrates the capability of our communication system to deeply analyze user instructions, accurately recognize the user's implicit intentions, and focus on the ROI for transmission. Taking Fig. \ref{fig:Pic3_intention} as an example, when watching and analyzing a football match, it is often desirable to quickly focus on the player who is controlling the ball. In this scenario, the instruction ``\emph{Please transmit the player who is controlling the ball}" is given to the communication system, enabling it to efficiently extract and transmit the relevant player from a large number of video frames. This task would be particularly challenging for traditional learning methods that don't utilize LMs. The semantic label (\emph{the player who is controlling the ball}) requires not only a strong mapping to the potential target area (\emph{the player}) but also an association between the target area and the specific context (\emph{who is controlling the ball}). In contrast, thanks to the powerful reasoning abilities of LMs, our system successfully performs the user-intention-driven communication task. Particularly in tasks involving the transmission of large volumes of images, our system can automatically extract all ROIs based on user instructions, significantly enhancing the overall task completion efficiency.

As shown in Fig. \ref{fig:4-metrics}, we use Peak Signal-to-Noise Ratio (PSNR, dB), Structural Similarity (SSIM), and perceptual similarity (LPIPS) based on a pre-trained VGG network as quantitative evaluation metrics to compare the performance of different methods under varying channel conditions. In the figure, under AWGN/Rayleigh channels, our method (including \textbf{Ours}, \textbf{w/o\ att$\mathrm{_{seg}}$}, and \textbf{w/o\ att$\mathrm{_{ori}}$}) demonstrates significant performance gains in PSNR, SSIM, and LPIPS, validating the effectiveness of the channel state awareness module. Additionally, as shown in Fig. \ref{fig:subfig1} and Fig. \ref{fig:subfig4}, \textbf{Ours} outperforms \textbf{w/o\ att$\mathrm{_{seg}}$} and \textbf{w/o\ att$\mathrm{_{ori}}$} in PSNR under all SNR conditions, validating the effectiveness of the MGA module. A similar trend is observed in the SSIM and LPIPS metrics. 
The numerical experimental results effectively validate the contributions of the channel awareness and MGA modules in enhancing the communication system's performance.

Fig. \ref{fig:Pic5_outputPicShow} visually demonstrates the advantages of our proposed method in terms of image transmission quality. As shown in the comparison, the transmission results of the baseline method show significant quality degradation compared to our method, particularly in detail-rich areas like text regions and the vehicle windshield. Our method demonstrates a superior ability to preserve texture details and maintain image clarity.


\section{Conclusion}
This paper proposes a User-Intention-Driven Semantic Communication system based on large models, overcoming the limitations of existing systems in understanding and generalizing users' real intentions. At the same time, by introducing a mask-guided attention module and a channel state awareness module, the system's transmission performance is significantly enhanced. The essential requirement of communication is to transmit the user's true intentions. We have equipped the communication system with the ability to deeply understand user transmission intentions, providing new ideas and methods for the further development of Semantic Communication.

\bibliographystyle{unsrt}
\bibliography{IEEEexample}



\end{document}